\documentclass[
    aps,         
    prc,         
    reprint,     
    twocolumn,  
    floatfix,    
    superscriptaddress 
]{revtex4-2}
\usepackage{amsmath} 
\usepackage{amssymb}
\usepackage{graphicx}
\usepackage{booktabs} 
\usepackage{orcidlink}
\usepackage{hyperref}
\begin{document}

\title{Measurement of the forward angle $^{12}\text{C} + ^{12}\text{C}$ fragmentation differential cross sections at $62~\text{MeV/nucleon}$}

\author{G. Guo}
\affiliation{School of Physics, Beihang University, Beijing 100191, China}

\author{G. Casini \orcidlink{0000-0001-8828-341X}}
\affiliation{INFN Sezione di Firenze, I-50019 Sesto Fiorentino, Italy}

\author{B. H. Sun \orcidlink{0000-0001-9868-5711}}
\email[Corresponding author, ]{bhsun@buaa.edu.cn}
\affiliation{School of Physics, Beihang University, Beijing 100191, China}

\author{S. Barlini \orcidlink{0000-0001-5991-2280}}
\affiliation{INFN Sezione di Firenze, I-50019 Sesto Fiorentino, Italy}
\affiliation{Dipartimento di Fisica, Università di Firenze, I-50019 Sesto Fiorentino, Italy}

\author{A. Camaiani \orcidlink{0000-0002-0417-0045}}
\affiliation{INFN Sezione di Firenze, I-50019 Sesto Fiorentino, Italy}

\author{C. Frosin \orcidlink{0000-0001-6314-7390}}
\affiliation{INFN Sezione di Firenze, I-50019 Sesto Fiorentino, Italy}
\affiliation{Dipartimento di Fisica, Università di Firenze, I-50019 Sesto Fiorentino, Italy}

\author{I. Lombardo}
\affiliation{INFN-Sezione di Catania, Catania, 95123, Italy}
\affiliation{Dipartimento di Fisica e Astronomia, Università di Catania, via S. Sofia 64, 95123 Catania, Italy}

\author{O. Lopez}
\affiliation{Université de Caen Normandie, ENSICAEN, CNRS/IN2P3, LPC Caen UMR6534, F-14000 Caen, France}

\author{S. Piantelli \orcidlink{0000-0002-0669-2787}}
\affiliation{INFN Sezione di Firenze, I-50019 Sesto Fiorentino, Italy}

\author{I. Tanihata}
\affiliation{School of Physics, Beihang University, Beijing 100191, China}
\affiliation{Research Center for Nuclear Physics (RCNP), The University of Osaka, Ibaraki, Osaka 567-0047, Japan}

\author{S. Terashima}
\affiliation{State Key Laboratory of Heavy Ion Science and Technology, Institute of Modern Physics, Chinese Academy of Sciences, Lanzhou 730000, China}

\author{S. Valdré \orcidlink{0000-0003-3629-6408}}
\affiliation{INFN Sezione di Firenze, I-50019 Sesto Fiorentino, Italy}

\author{G. Verde}
\affiliation{INFN-Sezione di Catania, Catania, 95123, Italy}

\author{F. W. Zhao}
\affiliation{School of Physics, Beihang University, Beijing 100191, China}

\author{L. Baldesi}
\affiliation{INFN Sezione di Firenze, I-50019 Sesto Fiorentino, Italy}
\affiliation{Dipartimento di Fisica, Università di Firenze, I-50019 Sesto Fiorentino, Italy}

\author{B. Borderie}
\affiliation{Université Paris-Saclay, CNRS/IN2P3, IJCLab, 91405 Orsay, France}

\author{R. Bougault \orcidlink{0000-0002-2866-7398}}
\affiliation{Université de Caen Normandie, ENSICAEN, CNRS/IN2P3, LPC Caen UMR6534, F-14000 Caen, France}

\author{C. Ciampi}
\affiliation{Grand Accélérateur National d'Ions Lourds (GANIL), CEA/DRF-CNRS/IN2P3, Boulevard Henri Becquerel, Caen, 14076, France}

\author{I. Dekhissi}
\affiliation{Université de Caen Normandie, ENSICAEN, CNRS/IN2P3, LPC Caen UMR6534, F-14000 Caen, France}

\author{J. A. Dueñas \orcidlink{0000-0003-2940-5135}}
\affiliation{Depto. de Ing. Eléctrica y Centro de Estudios Avanzados en Física, Matemáticas y Computación, Universidad de Huelva, S-21071 Huelva, Spain}

\author{Q. Fable}
\affiliation{Grand Accélérateur National d'Ions Lourds (GANIL), CEA/DRF-CNRS/IN2P3, Boulevard Henri Becquerel, Caen, 14076, France}
\affiliation{Laboratoire des 2 Infinis - Toulouse (L2IT-IN2P3), Université de Toulouse, CNRS, UPS, F-31062 Toulouse Cedex 9, France}

\author{F. Gramegna}
\affiliation{INFN-Laboratori Nazionali di Legnaro, Legnaro, 35020, Italy}

\author{D. Gruyer}
\affiliation{Université de Caen Normandie, ENSICAEN, CNRS/IN2P3, LPC Caen UMR6534, F-14000 Caen, France}

\author{A. Hocine \orcidlink{0009-0000-7767-9371}}
\affiliation{Université de Caen Normandie, ENSICAEN, CNRS/IN2P3, LPC Caen UMR6534, F-14000 Caen, France}

\author{B. Hong}
\affiliation{Department of Physics, Korea University, Seoul 02841, Republic of Korea}
\affiliation{Center for Extreme Nuclear Matters (CENuM), Korea University, 02841, Seoul, Republic of Korea}

\author{S. Kim}
\affiliation{Center for Exotic Nuclear Studies, Institute for Basic Science, Daejeon 34126, Republic of Korea}

\author{T. Kozik}
\affiliation{Marian Smoluchowski Institute of Physics, Jagiellonian University, Kraków, 30-348, Poland}

\author{S. H. Nam}
\affiliation{Department of Physics, Korea University, Seoul 02841, Republic of Korea}
\affiliation{Center for Extreme Nuclear Matters (CENuM), Korea University, 02841, Seoul, Republic of Korea}

\author{N. Le Neindre}
\affiliation{Université de Caen Normandie, ENSICAEN, CNRS/IN2P3, LPC Caen UMR6534, F-14000 Caen, France}

\author{J. Park}
\affiliation{Department of Physics, Korea University, Seoul 02841, Republic of Korea}
\affiliation{Center for Extreme Nuclear Matters (CENuM), Korea University, 02841, Seoul, Republic of Korea}

\author{M. Parlog}
\affiliation{Université de Caen Normandie, ENSICAEN, CNRS/IN2P3, LPC Caen UMR6534, F-14000 Caen, France}
\affiliation{``Horia Hulubei'' National Institute for R\&D in Physics and Nuclear Engineering (IFIN-HH), 077125 Bucharest Magurele, Romania}

\author{E. Vient}
\affiliation{Université de Caen Normandie, ENSICAEN, CNRS/IN2P3, LPC Caen UMR6534, F-14000 Caen, France}

\author{M. Vigilante}
\affiliation{Dipartimento di Fisica, Università di Napoli, 80126 Napoli, Italy}
\affiliation{INFN Sezione di Napoli, 80126 Napoli, Italy}

\author{C. J. Wang}
\affiliation{School of Physics, Beihang University, Beijing 100191, China}

\collaboration{FAZIA Collaboration}

\begin{abstract}
  The present work reports on high-precision measurements of forward-angle fragmentation differential cross sections for the $^{12}\text{C} + ^{12}\text{C}$ reaction at $62~\text{MeV/nucleon}$ using the FAZIA array. Angular distributions for fragments from $Z=1$ to $6$ were extracted in the range $2^{\circ} \leq \theta_{\text{lab}} \leq 8^{\circ}$. Particle identification was achieved by combining the $\Delta E\text{--}E$ technique with Pulse Shape Analysis, and precise energy calibrations were performed. 
  The results show that for heavier fragments, the angular distributions are better described by the Van Bibber formulation than by the Goldhaber model, consistent with the dominance of a wide component from dissipative processes in the measured angular range. Notably, $\alpha$ particles exhibit an anomalously narrow angular distribution, likely originating from the intrinsic cluster structure of $^{12}\text{C}$.
  Comparisons with existing data at the same incident energy show very good agreement, with a more complete set of species reported here, while a kinematic scaling is proposed to compare our data with previous experimental data at $50$ and $95~\text{MeV/nucleon}$.
\end{abstract}

\maketitle

\section{Introduction}\label{sec:introduction}
The complexity of nuclear reaction mechanisms in the Fermi energy regime has drawn significant interest in recent decades~\cite{Fatyga1985RPL}. This energy region serves as a transitional zone between direct fragmentation reactions, prevalent at higher energies, and compound-nucleus decay mechanisms, dominant at lower energies, thereby giving rise to complex reaction dynamics. To accurately characterize these dynamics, extensive measurements of heavy-ion reaction cross sections are essential for building a comprehensive database and rigorously testing theoretical models of these reactions~\cite{Fuchs1994RPP}.

Beyond their fundamental importance, such nuclear reaction data also have direct practical implications. In particular, carbon-ion radiotherapy (CIRT) has emerged as a prominent application that relies critically on precise knowledge of $^{12}\text{C}$ fragmentation cross sections. Recent advancements in accelerator technology have facilitated the global development of CIRT, which has been used to treat a large number of patients worldwide~\cite{web_PTCOG}, thereby motivating the need for high-precision $^{12}\text{C}$ reaction data across the therapeutic energy range~\cite{Mohamad2017Cancers}. Although initial beam energies in CIRT can reach up to $\sim 400~\text{MeV/nucleon}$, ions continuously slow down in biological tissue, eventually reaching the lower-energy regime of the present study as they approach the Bragg peak, where the major dose is delivered. Specifically, the nuclear fragmentation occurring during beam interaction in biological tissue must be accurately modeled to assess secondary particle damage, and positron-emitting fragments such as $^{11}\text{C}$ and $^{10}\text{C}$ offer a potential mechanism for range verification using PET, the accuracy of which depends on high-precision cross-section data~\cite{Kostyleva2023PMB}.

While a substantial body of measurements exists at higher incident energies ($\gtrsim 100~\text{MeV/nucleon}$)~\cite{Nandy2021FP}, the experimental database for the $^{12}\text{C} + ^{12}\text{C}$ system within the Fermi-energy regime ($\lesssim 100~\text{MeV/nucleon}$) remains poorly constrained. Data for this energy range are sparse at forward angles, leaving both forward-angle ($\theta_{\text{lab}} < 10^{\circ}$) differential cross sections and light-fragment yields insufficiently characterized~\cite{Dudouet2014PRC, Divay2017PRC,Napoli2012PMB}. This forward angular region is of paramount importance, as it accounts for a predominant portion of the total reaction yield. The lack of high-precision data in this dominant region thus represents a significant gap in our understanding of the complete fragmentation process at Fermi energies. The present study reports measurements of fragmentation differential cross sections for the $^{12}\text{C}$ + $^{12}\text{C}$ system at $62~\text{MeV/nucleon}$. Using the Forward mass-number ($A$) and charge-number ($Z$) Identification Array (FAZIA)~\cite{Bouga2014EPJA}, we provide high-precision data in the critical forward-angle region ($2^\circ \le \theta_{\text{lab}} \le 8^\circ$). 

This paper is organized as follows: Sec.~\ref{sec:setup} describes the experimental setup; Sec.~\ref{sec:data-analysis} details the data analysis procedures; Sec.~\ref{sec:discussion} presents the experimental results and a systematic comparison with literature data; and Sec.~\ref{sec:conclusion} provides a summary.

\section{Experimental Setup}\label{sec:setup}

\begin{figure*}[htbp]
\includegraphics
 [width=\textwidth]
 {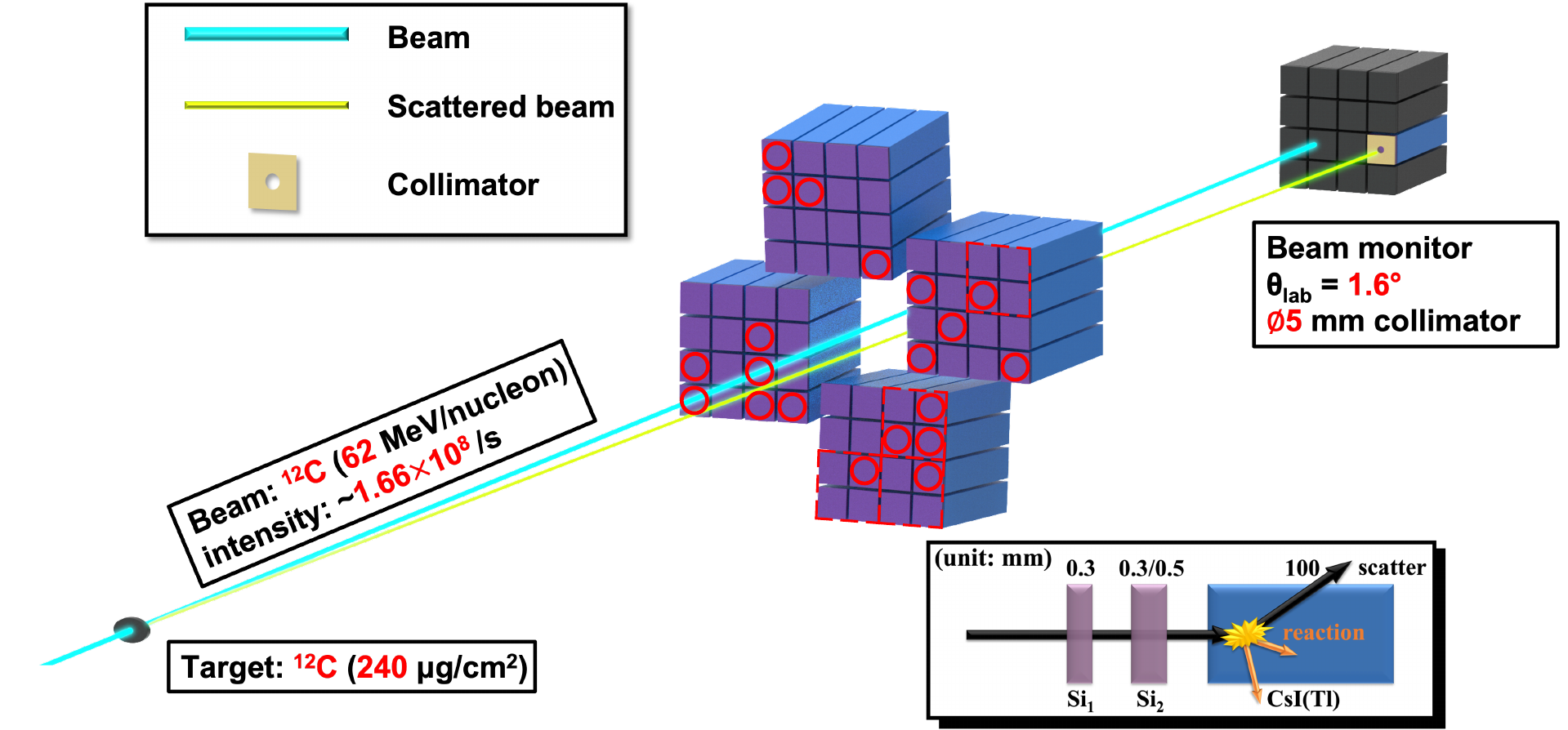}
 \caption{Schematic diagram of the experimental setup. The target and all detectors are housed within a vacuum chamber. The silicon detectors ($\text{Si}_1$ and $\text{Si}_2$) and $\text{CsI}(\text{Tl})$ detectors are colored purple and blue, respectively. Inactive detectors without readout are shown in black, and the collimator is shown in gold. Telescopes configured with a $300\,\mu\text{m}$ $\text{Si}_2$ layer (in groups of four) are outlined by red dashed boxes, while the rest are equipped with a $500\,\mu\text{m}$ $\text{Si}_2$ layer. The telescopes labeled with red open circles are excluded from the data analysis. The inset illustrates the structure of a basic telescope and includes a schematic representation of ion tracks in the CsI(Tl) crystal for the escaped particles.}
\label{fig:setup}
\end{figure*}

The experimental setup is illustrated in Fig.~\ref{fig:setup}. A $^{12}\text{C}$ beam at an energy of $62~\text{MeV/nucleon}$ was delivered by the superconducting cyclotron of the Istituto Nazionale di Fisica Nucleare - Laboratori Nazionali del Sud (INFN-LNS) and directed into a large vacuum chamber. With an intensity of approximately $1.66 \times 10^{8}~\text{s}^{-1}$, the beam was incident directly upon a $^{12}\text{C}$ target with a thickness of $240~\mu\text{g/cm}^2$. The target thickness is about $10^{-5}$ of the nuclear reaction mean free path and thus secondary effects within the target, such as multiple Coulomb scattering and cascade reactions, are negligible, allowing the study to focus on the fragmentation process itself.

The detection system was a subset of the FAZIA array positioned $1005~\text{mm}$ downstream of the target. It consisted of four block modules that covered polar angles from $2^\circ$ to $8^\circ$ to detect reaction products. Each block was equipped with 16 telescopes, and each telescope consists of a three-layer sequence: $\text{Si}_1$ with a thickness of $300~\mu\text{m}$, $\text{Si}_2$ with a thickness of either $300~\mu\text{m}$ (for 16 telescopes in total) or $500~\mu\text{m}$ (the remaining telescopes), and $\text{CsI(Tl)}$ with a thickness of $10~\text{cm}$, as illustrated in the inset of Fig.~\ref{fig:setup}. Each telescope has an angular acceptance of $\approx \pm 0.6^\circ$ in polar angle. The detectors were coupled to custom front-end electronics incorporating preamplifiers and fast digital sampling stages. Details of the FAZIA detector can be obtained from Refs.~\cite{Bouga2014EPJA, Valdre2019NIMA}.

  A separate monitor block was positioned $1814~\text{mm}$ downstream from the target. Within this block, only one telescope located at approximately $1.6^\circ$ was used for monitoring the beam intensity via the elastic scattering of $^{12}\text{C}$ ions. To define the solid angle, a $5~\text{mm}$ diameter gold collimator providing an angular acceptance of $\pm 0.16^\circ$ was placed in front of this telescope. For this reaction system, the classical laboratory grazing angle is estimated to be $0.55^\circ$, and the monitor angle is set to a slightly larger value. Since pure Rutherford normalization is no longer applicable, we use the optical potentials to evaluate the elastic cross section (see Sec.~\ref{sec:discussion}) and the uncertainties. Data acquisition was triggered by a threshold applied to a fast trapezoidal filter synthesized from the three detector signals (put in logic OR). Signal waveforms were recorded using digitization windows of $6~\mu\text{s}$ for the silicon detectors and $10~\mu\text{s}$ for the $\text{CsI(Tl)}$ detectors. Throughout the experiment, $\alpha$ sources ($^{239}\text{Pu}$, $^{241}\text{Am}$, and $^{243}\text{Cm}$) were placed inside the chamber to facilitate the energy calibration of the silicon detectors.

\section{Data Analysis}\label{sec:data-analysis}
\begin{table*}[htbp]
\centering
\caption{Energy thresholds of isotopes required to pass through $800\,\mu\text{m}$ silicon. Values are rounded to the nearest integer.}
\begin{tabular}{@{}l *{16}{c}@{}}
\toprule
{Isotope} & $^{1}\text{H}$ & $^{2}\text{H}$ & $^{3}\text{H}$ & $^{3}\text{He}$ & $^{4}\text{He}$ & $^{6}\text{Li}$ & $^{7}\text{Li}$ & $^{7}\text{Be}$ & $^{9}\text{Be}$ & $^{10}\text{Be}$ & $^{10}\text{B}$ & $^{11}\text{B}$ & $^{10}\text{C}$ & $^{11}\text{C}$ & $^{12}\text{C}$ & $^{13}\text{C}$\\
\midrule
$E_{\text{th}}\text{ (MeV)}$ & 11 & 14 & 17 & 38 & 43 & 81 & 86 & 120 & 133 & 139 & 180 & 187 & 221 & 230 & 239 & 247\\
 \bottomrule
  \end{tabular}
  \label{table:threshold}
\end{table*}
To extract reliable physical observables, the experimental data underwent several processing stages. First, the performance of all 64 telescopes was systematically evaluated. Twenty telescopes were excluded from the analysis due to either a lack of response in any detector layer or insufficient resolution for particle identification. Subsequently, a three-layer coincidence requirement was imposed within each telescope to suppress background events originating from scattering off structural materials. This criterion resulted in the rejection of approximately $4\%$ of the total recorded events, primarily affecting light fragments. As a result, only fragments that penetrate both silicon layers and reach the CsI(Tl) crystal are retained, which defines the effective energy thresholds for each isotope, as summarized in Table~\ref{table:threshold}. The tabulated values correspond to a typical telescope configuration with $300~\mu\text{m}$ $\text{Si}_1$ and $500~\mu\text{m}$ $\text{Si}_2$ detectors. For clarity, the subsequent description of the analysis procedure focuses on telescopes configured with $500~\mu\text{m}$ $\text{Si}_2$ detectors. Meanwhile, an identical methodology was applied to the $300~\mu\text{m}$ $\text{Si}_2$ subset.

\subsection{Particle Identification}\label{sec:pid}

\begin{figure}[htbp]
\includegraphics
 [width=\columnwidth]
 {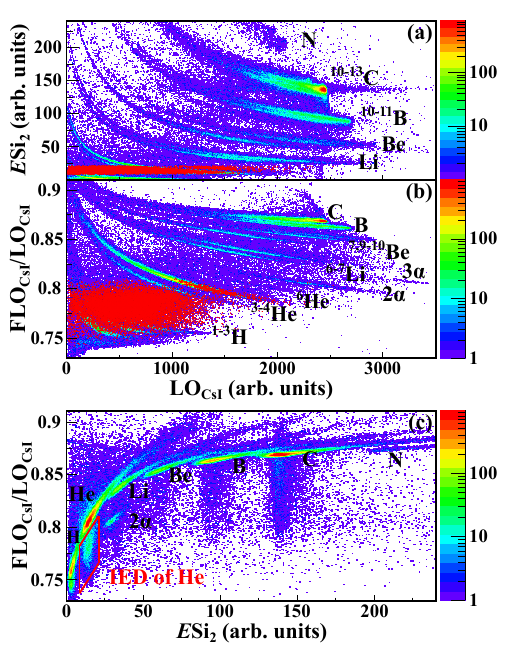}
 \caption{Particle identification methods in a representative telescope at $\theta_{\text{lab}} = 2.6^\circ$. (a) $\Delta E\text{--}E$ correlation; (b) PSA method; and (c) correlation between $\text{FLO}_{\text{CsI}}/\text{LO}_{\text{CsI}}$ and $E\text{Si}_2$. The red rectangular gate in (c) defines the selection region for IED events, which are then projected into the upper two panels as red scatter plots for comparison.}
 \label{fig:PID}
\end{figure}

Particle identification (PID) is achieved through two complementary methods: the $\Delta E\text{--}E$ technique and the pulse shape analysis (PSA) in the $\text{CsI(Tl)}$ scintillator. The selection between these two methods primarily depends on the particle energy and species, as illustrated in the experimental spectra of Figs.~\ref{fig:PID}(a) and (b). Specifically, the $\Delta E\text{--}E$ method correlates the energy loss in the $\text{Si}_2$ detector ($E\text{Si}_2$) with the total light output ($\text{LO}_{\text{CsI}}$), providing superior separation in the low-energy region. In contrast, the PSA method, which utilizes the ratio of the fast light output component ($\text{FLO}_{\text{CsI}}$) to $\text{LO}_{\text{CsI}}$, becomes more effective at higher energies. However, as evidenced by Fig.~\ref{fig:PID}(b), the PSA technique loses isotopic resolution for fragments with $Z \ge 5$. For these heavier fragments, the $\Delta E\text{--}E$ method remains the preferred identification tool across the entire energy range. The $E\text{Si}_1\text{--}E\text{Si}_2$ and PSA techniques in $\text{Si}_1$, although standard in other FAZIA experiments, are not performed for the analysis here, as this study focuses on ions that reach the $\text{CsI(Tl)}$ stage. They are used only for energy calibration, as discussed later.

A significant source of misidentification and particle inefficiency arises from the incomplete energy deposition (IED) events~\cite{Frosin2020NIMA,Baldesi2025NIMA} in the $\text{CsI(Tl)}$ detectors. These events occur when a particle fails to deposit its full energy within the scintillator volume, which can be caused by the particle scattering out of the crystal or by the production of secondary particles via nuclear reactions, as schematically illustrated in the inset of Fig.~\ref{fig:setup}. As a result, they appear in incorrect regions of the PID spectrum and are thus misidentified or lost. The presence of IED events may be clearly demonstrated by utilizing the scatter plot of $\text{FLO}_{\text{CsI}}/\text{LO}_{\text{CsI}}$ versus $E\text{Si}_2$, referred to as the $\text{PSA}\text{--}\Delta E$ spectrum and illustrated in Fig.~\ref{fig:PID}(c). The $\text{PSA}\text{--}\Delta E$ spectrum reveals distinct IED branches for each element (especially for light particles), which are located beneath the loci of regular events. This separation arises because IED events mostly originate from escaping particles, giving a smaller fast-light fraction due to reduced ionization density occurring when the Bragg peak is not fully contained in the crystal~\cite{Frosin2020NIMA}. For demonstration, a red parallelogram gate is drawn explicitly in Fig.~\ref{fig:PID}(c) to isolate the IED events of helium isotopes. Their impact is shown in Figs.~\ref{fig:PID}(a) and (b) with the superimposed red points, which clearly demonstrate how these events contaminate the identification regions of other isotopes and lead to their own misidentification when using either the $\Delta E\text{--}E$ or the PSA spectra alone. Although these IED events originate from genuine reaction products, misidentifying or losing them as incorrect particle types reduces detection efficiency, necessitating correction. In the following, efficiency corrections based on detailed simulations are applied. These simulations have been accurately benchmarked by our collaboration for protons~\cite{Frosin2020NIMA,Baldesi2025NIMA} and $\alpha$ particles~\cite{Baldesi2026NIMA}. A detailed discussion is provided in Sec.~\ref{sec:efficiency}.

 In events where multiple $\alpha$ particles hit the same CsI(Tl) detector simultaneously, PSA proves particularly useful. The constant $\text{FLO}_{\text{CsI}}/\text{LO}_{\text{CsI}}$ ratio for a given particle species allows for the identification of such multi-$\alpha$ events. As an example, in Fig.~\ref{fig:PID}(c), we label the region of double-$\alpha$ hits in the same crystal. This spot has the correct $\text{FLO}_{\text{CsI}}/\text{LO}_{\text{CsI}}$ ratio for $\alpha$ particles but approximately doubled energy, and thus falls within the IED region of lithium. In the following, such multi-$\alpha$ events are excluded from the $\alpha$ particle count.

 Overall, the final PID procedure consists of three stages. First, IED events are screened using the $\text{PSA}\text{--}\Delta E$ spectrum to ensure dataset purity. Second, the $\text{PSA}$ method is employed to identify $^{1,2,3}\text{H}$, $^{3,4}\text{He}$, $^{6,7}\text{Li}$, and $^{7,9,10}\text{Be}$ isotopes in the high-energy region. Third, the low-energy components of these light isotopes, along with the entire distributions of $^{10,11}\text{B}$ and $^{10,11,12,13}\text{C}$, are identified using the $\Delta E\text{--}E$ method. To efficiently process the large volume of experimental data, the identification of isotopes and the extraction of their yields were performed using the semi-automated routines in the KaliVeda toolkit~\cite{kaliveda}, as was done in all previous experiments with the FAZIA apparatus.

\subsection{Energy Calibration}

\begin{figure}[htbp]
\includegraphics
 [width=\columnwidth]
 {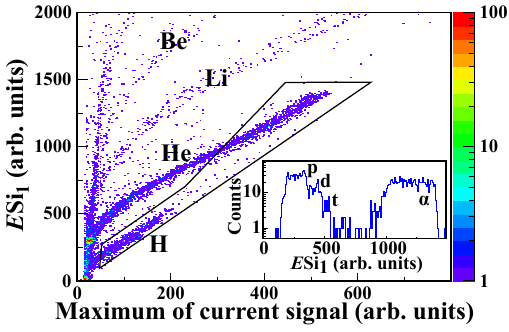}
 \caption{Correlation between the energy $E\text{Si}_1$ and the maximum of the current signal in the $\text{Si}_1$ detector for events stopped within the silicon stage. The projection onto the y-axis of the region enclosed by the black box is shown in the inset. This figure depicts a representative telescope located at $\theta_{\text{lab}} = 4.2^\circ$.}
\label{fig:sicali_PSA}
\end{figure}

\begin{figure}[htbp]
\includegraphics
 [width=\columnwidth]
 {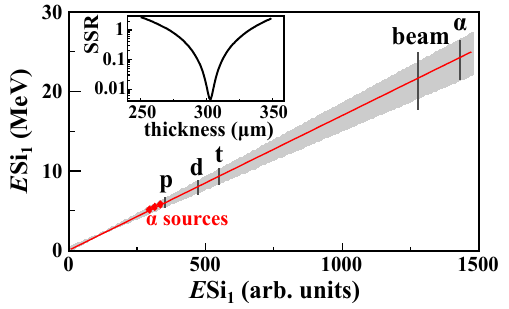}
 \caption{An example of the energy calibration results for the $\text{Si}_1$ detector. The calibrated energy loss $E\text{Si}_1$ ($\text{MeV}$) is plotted against the raw energy signal in channels. The three red points represent the signals from $\alpha$ sources, while the vertical bars indicate the thickness-dependent energy reference marks for PTPs and beam particles. The solid red line shows the best fit. The inset displays the Sum of Squared Residuals (SSR) value for the linear energy calibration as a function of the assumed thickness.}
\label{fig:cali_si1}
\end{figure}

\begin{figure}[htbp]
\includegraphics
 [width=\columnwidth]
 {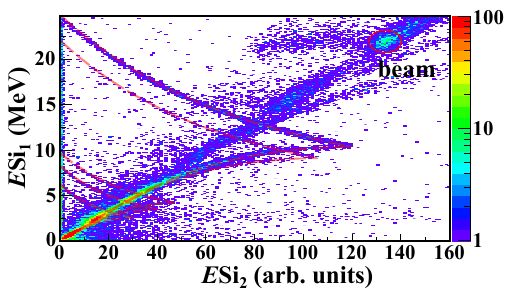}
\caption{Energy loss correlation between the two silicon detectors, plotting the calibrated energy loss $E\text{Si}_1$ (MeV) versus the raw energy loss $E\text{Si}_2$ (a.u.). The dashed red lines are derived from energy loss tables, calculated post-calibration to verify the experimental loci. The intense spot highlighted by the red ellipse in the upper-right corner corresponds to the elastic beam particles.}
\label{fig:sicali_dEdE}
\end{figure}

The energy calibration is performed sequentially from the front to the back detectors of the telescope, using the energy information from each calibrated preceding stage to calibrate the subsequent one. For the $\text{Si}_1$ detectors, although $\alpha$ sources ($^{239}\text{Pu}$, $^{241}\text{Am}$, and $^{243}\text{Cm}$) were recorded by the $\text{Si}_1$ detectors, their limited energy range could lead to significant extrapolation errors at high energies.

To provide a more robust calibration across the full dynamical range, we incorporated ions from the elastically scattered beam and several punch-through points (PTPs). The beam calibration point was derived from events where the primary beam particles, elastically scattered from the target, stopped in the subsequent $\text{CsI}(\text{Tl})$ crystal after traversing the silicon layers (see the red marked spot in Fig.~\ref{fig:sicali_dEdE}). Although particles stopping in the silicon layers are excluded from the final physics yield, they remain indispensable for determining these PTPs. Taking one telescope as a representative example, Fig.~\ref{fig:sicali_PSA} displays the correlation between the energy deposited in $\text{Si}_1$ and the maximum current signal, a typical PSA parameter used by the FAZIA collaboration~\cite{Pastore2017NIMA}. Distinct elemental loci are clearly resolved, representing particles that stop in the first silicon layer. By applying specific graphical gates to each element, the PTPs for various isotopes can be precisely identified at the boundary of each locus, where particles begin to enter the $\text{Si}_2$ detector. The identification process is illustrated in the inset of Fig.~\ref{fig:sicali_PSA}, which shows the projection onto the $E\text{Si}_1$ axis for events within the black box. The inset reveals the characteristic ``cliff-edge'' structures used to define the PTPs, with each particle type ($p$, $d$, $t$, and $\alpha$) labeled accordingly.

A critical aspect of the calibration process is determining the exact active thickness of the detector. Unlike the $\alpha$ source peaks, which are independent of the detector thickness due to their short range, the energy loss of elastically scattered beam particles and PTPs is highly sensitive to the silicon's active depth. Consequently, rather than relying on the nominal thickness, we treated the detector thickness as a free parameter. As shown in Fig.~\ref{fig:cali_si1}, the calculated energy losses for the beam particles and PTPs vary significantly with the assumed thickness (represented by vertical bars), whereas the $\alpha$ source data remain constant. To determine the optimal value, we performed a series of linear calibrations by scanning the assumed thickness from $250~\mu\text{m}$ to $350~\mu\text{m}$ in $1~\mu\text{m}$ increments.
The optimal thickness and the related calibration coefficients were determined by minimizing the Sum of Squared Residuals (SSR) value of the linear fit (see inset of Fig.~\ref{fig:cali_si1}). The resulting effective thicknesses for all $\text{Si}_1$ detectors span a range of $290\text{--}320~\mu\text{m}$. After completion of $\text{Si}_1$ calibration, the $\text{Si}_2$ detectors were calibrated using the $E\text{Si}_1$ vs. $E\text{Si}_2$ correlation depicted in Fig.~\ref{fig:sicali_dEdE}, which includes both particles stopping in $\text{Si}_2$ and those penetrating into the $\text{CsI}(\text{Tl})$ crystal. The calibration coefficients for $\text{Si}_2$ were adjusted to align the experimental ridges with theoretical isotope-specific energy-loss curves, after which we attempted to match the theoretical PTPs to the experimental cusps by further tuning the detector thickness. An optimal thickness was determined in $1~\mu\text{m}$ steps by minimizing the overall discrepancies among the PTPs~\cite{Alex2024thesis} for fragments ranging from $Z=1$ to $Z=3$. The resulting value fell within approximately $\pm 30~\mu\text{m}$ of the nominal thickness. PTPs for $Z>3$ were not clearly discernible and thus excluded from the optimization.

For the $\text{CsI(Tl)}$ detectors, the energy deposited by $Z>2$ nuclides was deduced from the energy loss in the preceding silicon layers, which yields an acceptable resolution of less than $10\%$. However, for hydrogen and helium isotopes, the typical energy loss in the silicon detectors is too small to provide a precise correlation, resulting in a deduced energy resolution in $\text{CsI(Tl)}$ that exceeds $20\%$. To address this, we performed direct energy calibrations for the $Z=1$ and $Z=2$ isotopes using the low-energy region, where the silicon energy loss is sufficiently large to deduce a reliable residual energy. A global calibration was performed for these light particles using a single functional form that accounts for the quenching effect~\cite{Parlog2002NIMA} as a function of the charge number $Z$ and mass number $A$.

\subsection{Identification Efficiency}\label{sec:efficiency}

\begin{figure}[htbp]
\includegraphics
 [width=\columnwidth]
 {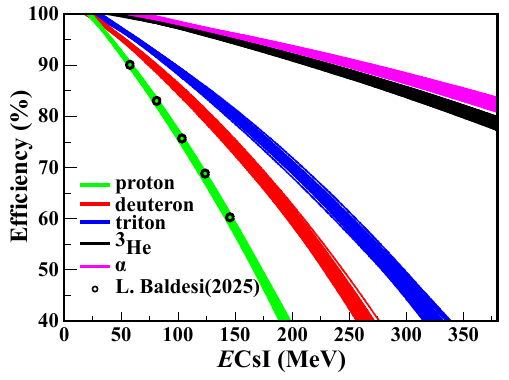}
 \caption{The simulated identification efficiency for nuclides with $Z \leq 2$ in the CsI(Tl) detectors. The solid lines represent the simulated results, with colors differentiating the various nuclides. Each shaded band indicates the spread arising from the overlap of 64 individual crystal responses. Data points represent results from previous measurements for comparison.}
\label{fig:efficiency}
\end{figure}

Following the energy calibration, we obtained the residual energy of each particle as it enters the $\text{CsI(Tl)}$ crystal. This information enables a simulation-based determination of the particle-identification efficiency in the CsI(Tl) scintillator, which is essential for correcting IED events. The Geant4 simulation toolkit was utilized to determine the efficiency. The simulation employed the QBBC physics list, with G4EmStandardPhysics\_option4 used to optimize the low-energy electromagnetic interactions~\cite{morfo2017NIMA,Swean2021NIMA}. Initial particle positions were sampled on the downstream target surface according to a Gaussian beam-spot profile with $\sigma = 2~\text{mm}$, and the particles were then emitted according to the measured angular distributions for each specific isotope. The simulated efficiency is defined as the ratio of the full-energy peak counts to the total counts recorded in the $\text{CsI(Tl)}$ energy spectrum. Through systematic simulations at varying incident energies, the efficiency was parameterized as a quadratic function of the total deposited energy within the CsI(Tl) crystals. The results are presented in Fig.~\ref{fig:efficiency}, with colors differentiating the various nuclides. Each shaded band arises from the overlap of 64 individual crystal responses. In the following data analysis, the efficiency correction is applied to every active telescope. This simulation framework reproduces the proton efficiency as measured in a previous dedicated experiment by our collaboration~\cite{Baldesi2025NIMA}, validating its reliability. Under the present experimental conditions, the measured counts for protons, deuterons, tritons, $^{3}\text{He}$, and $\alpha$ particles are multiplied by factors of approximately 1.10, 1.15, 1.20, 1.05, and 1.08, respectively, to account for energy-dependent detection inefficiencies. For heavier fragments, no correction is applied, i.e., full efficiency is assumed.

\section{Results and Discussion}\label{sec:discussion}

The absolute scale of the differential cross sections was determined by calibrating the fragment yields against the beam-elastic-scattering events. The calculation is expressed as
\begin{equation}
  \frac{d\sigma}{d\Omega}({}^{A}_{Z}\text{X})=\sigma_{\text{el}}\frac{N(^{A}_{Z}\text{X})}{N(^{12}\text{C})}\frac{1}{\Omega} \;,
 \label{eq:fcs}
\end{equation}
where $\sigma_{\text{el}}$ is the theoretical elastic scattering cross section for the monitor integrated over its solid angle, $N(^{12}\text{C})$ is the count of beam $^{12}\text{C}$ particles elastically scattered into the monitor telescope, $N(^{A}_{Z}\text{X})$ is the number of fragments with mass $A$ and charge $Z$, and $\Omega$ is the solid angle of each telescope.

The cross section $\sigma_{\text{el}}$ was calculated using the FRESCO code~\cite{Thompson1988CPR} with the optical model potential parameters taken from Ref.~\cite{Buenerd1984NPA}. An uncertainty of 10\% was assigned to $\sigma_{\text{el}}$ to account for the discrepancies between different parameter sets. 
The solid angle $\Omega$ has an estimated geometric uncertainty of 1\%.
The uncertainty in the fragment count $N$ includes the statistical error for all fragments. For $Z=1$ and $Z=2$ isotopes, systematic uncertainties could arise from the efficiency calculation. However, as discussed above, the simulation framework has been benchmarked against dedicated experiments for protons and alpha particles. Given this validation, the systematic uncertainties are considered negligible.

\begin{figure}[htbp]
\includegraphics
 [width=\columnwidth]
 {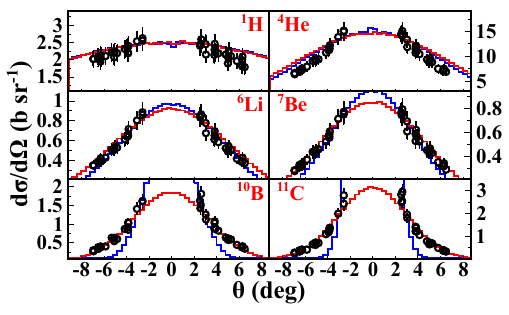}
\caption{Differential cross section angular distributions for various outgoing nuclides. Each data point represents the experimental cross section measured by an individual telescope. The blue and red lines correspond to calculations using Goldhaber and Van Bibber formulations, respectively.}
\label{fig:trend}
\end{figure}

\begin{figure}[htbp]
\includegraphics
 [width=\columnwidth]
 {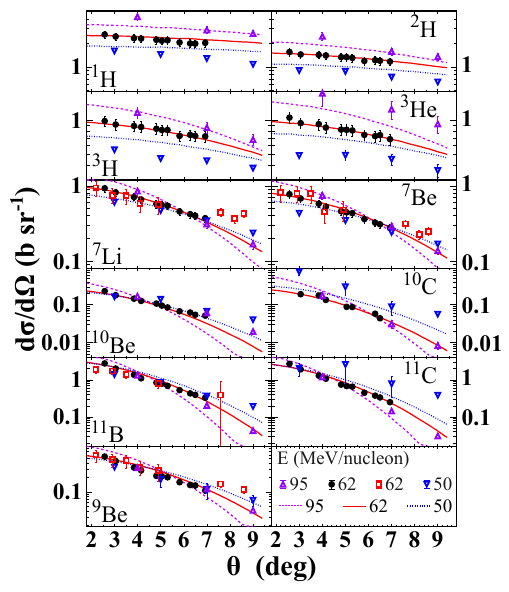}
 \caption{Comparison of the measured angular differential cross sections with results from previous studies~\cite{Dudouet2014PRC,Napoli2012PMB,Divay2017PRC}. Points represent the experimental data, and curves indicate the distributions calculated using the Van Bibber formulation with a constant intrinsic momentum width $W$ to illustrate the expected kinematic scaling.}
\label{fig:other}
\end{figure}

The inclusive fragmentation differential cross sections are presented in Fig.~\ref{fig:trend} as a function of the central polar angle $\theta_{\text{lab}}$ of the telescopes. Positive and negative values of $\theta_{\text{lab}}$ correspond to detectors on the $x > 0$ and $x < 0$ sides, respectively. In total, sixteen nuclides were measured, and representative angular distributions for one selected isotope of each element are shown in Fig.~\ref{fig:trend} and distributions for other fragments are shown in Fig.~\ref{fig:other}.

All observed fragments exhibit clear symmetry around $\theta_{\text{lab}}=0^\circ$, which confirms that the cross sections measured by telescopes at equivalent polar angles are mutually consistent. This agreement is particularly significant because each telescope was analyzed independently, demonstrating that the calibrations and corrections applied to the data are consistent and reliable. This symmetry further reflects the expected azimuthal invariance of the reaction process around the beam axis. Consequently, data from telescopes at equivalent scattering angles $|\theta_{\text{lab}}|$ were averaged to determine the final cross section for each angular position. In rare instances where counts for specific nuclides showed discrepancies between telescopes at the same polar angle that exceeded statistical expectations, these variations were primarily attributed to localized differences in detector identification resolution. Nevertheless, to ensure a conservative estimation, such discrepancies were carefully incorporated into the systematic uncertainty. The resulting cross sections, including both statistical and systematic contributions, are compiled in Table~\ref{table:results}.

\begin{table*}[htbp]
\centering
\caption{Angular distribution of the differential cross sections $d\sigma/d\Omega$ for the $^{12}\text{C} + ^{12}\text{C}$ reaction at $62~\text{MeV/nucleon}$. Results from telescopes at equivalent $|\theta_{\text{lab}}|$ have been averaged to improve statistical precision. Uncertainties represent the sum of statistical and systematic contributions.}
    \begin{tabular}{lcccccccc}
        \toprule
        $\theta_{\text{lab}}$ (deg) & $^{1}\text{H}$ (b sr$^{-1}$) & $^{2}\text{H}$ (b sr$^{-1}$)& $^{3}\text{H}$ (b sr$^{-1}$)& $^{3}\text{He}$ (b sr$^{-1}$)& $^{4}\text{He}$ (b sr$^{-1}$)& $^{6}\text{Li}$ (b sr$^{-1}$)& $^{7}\text{Li}$ (b sr$^{-1}$)& $^{7}\text{Be}$ (b sr$^{-1}$)\\
        \midrule
        2.6(0.7) & 2.5(0.3) & 1.54(0.19) & 0.98(0.12) & 1.05(0.12) & 14.9(1.6) & 0.87(0.09) & 0.93(0.10) & 0.78(0.08) \\
        3.1(0.8) & 2.40(0.29) & 1.43(0.18) & 0.92(0.11) & 0.94(0.10) & 13.1(1.4) & 0.76(0.08) & 0.82(0.09) & 0.68(0.07) \\
        3.8(0.7) & 2.29(0.28) & 1.41(0.18) & 0.89(0.11) & 0.92(0.10) & 11.5(1.3) & 0.66(0.07) & 0.71(0.08) & 0.58(0.06) \\
        4.2(0.8) & 2.26(0.28) & 1.38(0.17) & 0.88(0.11) & 0.86(0.10)  & 10.6(1.2) & 0.60(0.07) & 0.66(0.07) & 0.53(0.06) \\
        4.8(0.8) & 2.18(0.27) & 1.33(0.16) & 0.84(0.10)  & 0.83(0.09)  & 9.4(1.0)  & 0.53(0.06) & 0.57(0.06) & 0.46(0.05) \\
        5.1(0.6) & 2.12(0.26) & 1.32(0.16) & 0.82(0.10)  & 0.82(0.09)  & 9.2(1.0)  & 0.51(0.06) & 0.56(0.06) & 0.45(0.05) \\
        5.3(0.7) & 2.16(0.27) & 1.29(0.16) & 0.82(0.10)  & 0.81(0.09)  & 9.0(1.0)  & 0.51(0.06) & 0.54(0.06) & 0.43(0.05) \\
        5.8(0.8) & 2.03(0.25) & 1.18(0.15) & 0.73(0.09)  & 0.74(0.08)  & 7.7(0.9)   & 0.42(0.05) & 0.46(0.05) & 0.36(0.04) \\
        6.3(0.6) & 1.97(0.24) & 1.20(0.15) & 0.74(0.09)  & 0.72(0.08)  & 7.2(0.8)   & 0.40(0.04) & 0.42(0.05) & 0.32(0.04) \\
        6.5(0.7) & 1.95(0.24) & 1.19(0.15) & 0.73(0.09)  & 0.74(0.08)  & 7.0(0.8)   & 0.38(0.04) & 0.40(0.04) & 0.31(0.03) \\
        6.9(0.7) & 2.00(0.25) & 1.17(0.15) & 0.72(0.09)  & 0.68(0.08)  & 6.5(0.7)   & 0.34(0.04) & 0.37(0.04) & 0.28(0.03) \\
        \midrule
        \addlinespace
        $\theta_{\text{lab}}$ (deg) & $^{9}\text{Be}$ (b sr$^{-1}$)& $^{10}\text{Be}$ (b sr$^{-1}$)& $^{10}\text{B}$ (b sr$^{-1}$)& $^{11}\text{B}$ (b sr$^{-1}$)& $^{10}\text{C}$ (b sr$^{-1}$)& $^{11}\text{C}$ (b sr$^{-1}$)& $^{12}\text{C}$ (b sr$^{-1}$)& $^{13}\text{C}$ (b sr$^{-1}$)\\
        \midrule
        2.6(0.7) & 0.55(0.06) & 0.231(0.025) & 1.62(0.18) & 2.8(0.3) & -- & 2.70(0.29) & 15.1(1.6) & 0.67(0.09) \\
        3.1(0.8) & 0.45(0.05) & 0.180(0.020) & 1.27(0.14) & 2.02(0.22) & 0.190(0.023) & 1.93(0.21) & 8.3(1.2) & -- \\
        3.8(0.7) & 0.32(0.04) & 0.145(0.016) & 0.96(0.10) & 1.37(0.15) & 0.175(0.023) & 1.30(0.14) & 3.7(0.4) & 0.213(0.023) \\
        4.2(0.8) & 0.28(0.03) & 0.127(0.014) & 0.85(0.09) & 1.10(0.12) & 0.135(0.015) & 1.08(0.12) & 3.0(0.3) & 0.176(0.019) \\
        4.8(0.8) & 0.218(0.024) & 0.107(0.012) & 0.62(0.07) & 0.85(0.09) & -- & 0.74(0.08) & 1.84(0.20) & 0.129(0.017) \\
        5.1(0.6) & 0.214(0.023) & 0.095(0.010) & 0.59(0.06) & 0.78(0.08) & 0.088(0.011) & 0.68(0.07) & 1.70(0.18) & 0.116(0.017) \\
        5.3(0.7) & 0.202(0.022) & 0.085(0.009) & 0.61(0.07) & 0.69(0.08) & 0.087(0.010) & 0.66(0.07) & 1.40(0.15) & -- \\
        5.8(0.8) & 0.161(0.018) & 0.067(0.007) & 0.42(0.05) & 0.53(0.06) & -- & 0.45(0.05) & 1.01(0.11) & 0.065(0.009) \\
        6.3(0.6) & 0.140(0.015) & 0.062(0.007) & 0.38(0.04) & 0.43(0.05) & 0.058(0.006) & 0.38(0.04) & 0.73(0.08) & 0.050(0.006) \\
        6.5(0.7) & 0.137(0.015) & 0.055(0.006) & 0.34(0.04) & 0.39(0.04) & 0.044(0.005) & 0.34(0.04) & 0.67(0.08) & 0.037(0.004) \\
        6.9(0.7) & 0.107(0.012) & 0.052(0.006) & 0.28(0.03) & 0.33(0.04) & -- & 0.251(0.027) & 0.55(0.06) & -- \\
        \midrule
    \end{tabular}
\label{table:results}
\end{table*}

A systematic narrowing of the angular distributions with increasing fragment mass $A$ is observed in our experimental data. This trend indicates that the transverse momentum distribution is sensitive to the fragment species. To reproduce these distributions, we employed an empirical momentum-space model where fragments are assumed to inherit the projectile's velocity, with their longitudinal momentum widths $W_L$ governed by the Goldhaber statistical model~\cite{Goldhaber1974PLB}:

\begin{equation}
  W_L^2 = W_0^2 \frac{A_f(A_p - A_f)}{A_p - 1}.
\end{equation}

Here, $A_f$ and $A_p$ are the mass numbers of the fragments and the projectile, respectively. The intrinsic nucleon momentum spread $W_0$ is set to $95~\text{MeV}/c$~\cite{Giacomelli2004PRC}. For the transverse momentum $p_{\perp}$, we compared the isotropic Goldhaber assumption ($W_{\perp} = W_L$) with an extended formulation proposed by Van Bibber \textit{et al.}~\cite{VanBibber1979PRL}:

\begin{equation}
  W_{\perp}^2 = W_1^2 \frac{A_f(A_p - A_f)}{A_p - 1} + W_2^2 \frac{A_f(A_f - 1)}{A_p(A_p - 1)}.
\end{equation}

In this approach, $W_1$ is set to be the same as the longitudinal width ($95~\text{MeV}/c$), while an additional transverse width $W_2$ is introduced to account for the deflection effect originating from both Coulomb repulsive and nuclear attractive forces~\cite{VanBibber1979PRL}. Its value depends on incident energy and target nuclear charge. For our carbon target, we adopt $W_2 = 195~\text{MeV}/c$~\cite{Notani2007PRC,Momota2023PhysScr}. 
Within the small angular acceptance, no $Z$-dependence is observed~\cite{wong1982PRC}.

Theoretical angular distributions were generated via Monte Carlo sampling and normalized to the experimental data at $\theta_{\text{lab}} = 3^\circ$ to focus on the comparison of distribution shapes. As shown in Fig.~\ref{fig:trend}, while both models align for light fragments, a clear divergence appears for heavier species like $^{10}\text{B}$ and $^{11}\text{C}$, where the inclusion of the $W_2$ term provides a superior fit. Interestingly, for these heavy fragments, a slight enhancement in the cross section is observed around $2.5^\circ$, where the data tend to favor the narrower profile aligned with the Goldhaber model. Such narrow components are also reported in Ref.~\cite{Momota2023PhysScr}. The Van Bibber model, however, predicts only one wide component, assuming that the distribution widens due to orbital deflection. If this deflection process were dominant, all fragments would be bent and show only a wide component. In contrast, if the widening arises from the de-excitation of pre-fragments, unexcited direct fragments can retain a narrow angular distribution, thereby naturally explaining the observed two-component structure. Indeed, deflection calculations show that this effect is negligibly small for a light target such as carbon~\cite{Momota2023PhysScr}. Therefore, the current angular distributions of fragments indicate that direct fragmentation and subsequent de-excitation of pre-fragments are the dominant processes at the present beam energy for this reaction system.

Notably, for $\alpha$ particles, the experimental distribution remains significantly narrower than the predictions of both models, consistent with the observations by Greiner \textit{et al.}~\cite{Greiner1975PRL}. This discrepancy may arise primarily from the intrinsic cluster structure of the $\alpha$ particle; as a pre-formed and strongly bound entity, its emission during fragmentation avoids the random momentum summation of independent nucleons, effectively ``freezing'' the internal degrees of freedom and leading to a contraction in momentum space.

As illustrated by the various symbol shapes in Fig.~\ref{fig:other}, our results were benchmarked against three other studies~\cite{Dudouet2014PRC, Napoli2012PMB, Divay2017PRC} using the same reaction system at similar incident energies. At the same incident energy of $62~\text{MeV/nucleon}$, our data are consistent with previous results within experimental uncertainties, while providing cross sections for a wider range of fragment species and achieving improved precision across all measured distributions. Comparing across the different incident energies, we find that the larger the incident beam energy, the larger the differential cross section for light fragments from $^{1}\text{H}$ to $^{3}\text{He}$, but such a pattern is not observed for fragments heavier than $^{3}\text{He}$.

To elucidate the mechanism behind this trend, we must first distinguish it from a trivial kinematic effect arising from the change in beam velocity, which could obscure the physics of interest. For this purpose, we adopt the width $W_{\perp}$ based on the Van Bibber formulation simply because it provides a good fit to the present data, as shown in Fig.~\ref{fig:trend}. However, the deflection mechanism is not invoked. Instead, we attribute the observed broadening to the de-excitation of pre-fragments and consequently assume that $W_{\perp}$ does not depend on beam energy. The curves for other incident energies displayed in Fig.~\ref{fig:other} are obtained by fixing the parameters of the momentum width $W_1$ and $W_2$ to the value used at $62~\text{MeV/nucleon}$, varying only the beam velocity according to the incident energy, and applying the same normalization factor as for the 62 MeV/nucleon data. Under this procedure, the integrated fragmentation cross section is kept constant across all energies, meaning that any remaining differences between the data and the scaled curves would reflect genuine changes in the underlying reaction mechanism beyond simple kinematic scaling. 

At $95~\text{MeV/nucleon}$, the scaled theoretical distributions successfully reproduce the measured distributions for light isotopes ($^{1,2,3}\text{H}$). For $^3\text{He}$, however, the experimental values are higher than the predicted curve. This discrepancy may arise from residual uncertainties in the correction for $\alpha$ misidentification applied to the $^3$He data~\cite{Dudouet2013PRC}, or reflect the opening of additional reaction channels for $^3$He at this energy. For heavier fragments, although the theoretical curves align with the data at smaller angles, a slightly broader component remains in the experimental cross sections.

In contrast, at $50~\text{MeV/nucleon}$, the theoretical calculations generally overestimate the yields for light nuclei while underestimating the heavier fragments. This discrepancy suggests that at this lower energy, the de-excitation (decay) of primary fragments is reduced compared to the $62~\text{MeV/nucleon}$ case. Consequently, heavier fragments survive rather than breaking apart, leading to experimental yields that exceed theoretical predictions. Conversely, the production of light fragments from these decay chains is reduced, leading to the model's observed overestimation. However, it appears to deviate from the usual expectation that direct fragmentation becomes more dominant at higher energies, while de-excitation processes prevail at lower energies. Further studies are needed to clarify this behavior.

\section{Conclusion}\label{sec:conclusion}
In this work, we have performed high-precision measurements of forward-angle fragmentation differential cross sections for the $^{12}\text{C} + ^{12}\text{C}$ reaction at $62~\text{MeV/nucleon}$ using the FAZIA array. Angular distributions for sixteen isotopes were extracted in the range $2^{\circ} \leq \theta_{\text{lab}} \leq 8^{\circ}$. Reliable particle identification was achieved by combining the $\Delta E\text{--}E$ technique and PSA method. A self-consistent energy calibration was implemented by treating the detector active thicknesses as free parameters. Efficiency losses due to incomplete energy deposition in the active volume were corrected via dedicated Geant4 simulations.

The analysis of the angular distributions over the measured range shows that for projectile-like fragments, the Van Bibber formulation of the transverse momentum reproduces the experimental results better than the standard Goldhaber model, while both models yield comparable results for the lightest species ($Z=1, 2$). Notably, $\alpha$ particles exhibit an unexpectedly narrow distribution, which may be attributed to the intrinsic cluster structure of $^{12}\text{C}$. Enhanced cross sections at small scattering angles are seen for heavy fragments, showing that the direct fragmentation process still contributes. This is consistent with the negligible deflection of outgoing fragments by nuclear potential and Coulomb interaction at this energy with a light target. 

Our results are in full agreement with published data at $62~\text{MeV/nucleon}$, while providing cross sections for a broader set of isotopes with improved precision. Moreover, systematic comparisons with experimental data at $50$ and $95~\text{MeV/nucleon}$ demonstrate that the cross sections from $62$ to $95~\text{MeV/nucleon}$ are well reproduced by kinematic scaling. In contrast, this scaling approach fails to describe the results at $50~\text{MeV/nucleon}$, where significant mass-dependent deviations emerge: the experimental yields for light fragments are lower than the scaling predictions, while those for heavier fragments are higher. It suggests that the de-excitation during the fragmentation process is sensitive to the incident energy. Specifically, at lower energies, the decay process may be weaker, leading to a higher survival probability of primary heavy fragments. These findings provide new, precise experimental data to understand the fragmentation of light nuclei, particularly $\alpha$-cluster nuclei, at forward angles. A systematic comparison with transport-model calculations is being prepared to further elucidate the dynamical mechanisms underlying the fragmentation of light nuclear systems.

\begin{acknowledgments}
This work was supported by the National Natural Science Foundation of China (Nos. 12325506, 12550008, 12175009), the Institute for Basic Science (IBS-R031-D1) in Korea, and the National Research Foundation of Korea (NRF) (Grant Nos. 2018R1A5A1025563 and RS-2024-00333673). The whole FAZIA Collaboration thanks the INFN-LNS beam operators and the target manufacturing laboratory for the quality of the delivered beams and the produced targets. G. Guo gratefully acknowledges financial support from the China Scholarship Council.
\end{acknowledgments}


\begin{thebibliography}{30}%
\makeatletter
\providecommand \@ifxundefined [1]{%
 \@ifx{#1\undefined}
}%
\providecommand \@ifnum [1]{%
 \ifnum #1\expandafter \@firstoftwo
 \else \expandafter \@secondoftwo
 \fi
}%
\providecommand \@ifx [1]{%
 \ifx #1\expandafter \@firstoftwo
 \else \expandafter \@secondoftwo
 \fi
}%
\providecommand \natexlab [1]{#1}%
\providecommand \enquote  [1]{``#1''}%
\providecommand \bibnamefont  [1]{#1}%
\providecommand \bibfnamefont [1]{#1}%
\providecommand \citenamefont [1]{#1}%
\providecommand \href@noop [0]{\@secondoftwo}%
\providecommand \href [0]{\begingroup \@sanitize@url \@href}%
\providecommand \@href[1]{\@@startlink{#1}\@@href}%
\providecommand \@@href[1]{\endgroup#1\@@endlink}%
\providecommand \@sanitize@url [0]{\catcode `\\12\catcode `\$12\catcode `\&12\catcode `\#12\catcode `\^12\catcode `\_12\catcode `\%12\relax}%
\providecommand \@@startlink[1]{}%
\providecommand \@@endlink[0]{}%
\providecommand \url  [0]{\begingroup\@sanitize@url \@url }%
\providecommand \@url [1]{\endgroup\@href {#1}{\urlprefix }}%
\providecommand \urlprefix  [0]{URL }%
\providecommand \Eprint [0]{\href }%
\providecommand \doibase [0]{https://doi.org/}%
\providecommand \selectlanguage [0]{\@gobble}%
\providecommand \bibinfo  [0]{\@secondoftwo}%
\providecommand \bibfield  [0]{\@secondoftwo}%
\providecommand \translation [1]{[#1]}%
\providecommand \BibitemOpen [0]{}%
\providecommand \bibitemStop [0]{}%
\providecommand \bibitemNoStop [0]{.\EOS\space}%
\providecommand \EOS [0]{\spacefactor3000\relax}%
\providecommand \BibitemShut  [1]{\csname bibitem#1\endcsname}%
\let\auto@bib@innerbib\@empty
\bibitem [{\citenamefont {Fatyga}\ \emph {et~al.}(1985)\citenamefont {Fatyga}, \citenamefont {Kwiatkowski}, \citenamefont {Viola}, \citenamefont {Chitwood}, \citenamefont {Fields}, \citenamefont {Gelbke}, \citenamefont {Lynch}, \citenamefont {Pochodzalla}, \citenamefont {Tsang},\ and\ \citenamefont {Blann}}]{Fatyga1985RPL}%
  \BibitemOpen
  \bibfield  {author} {\bibinfo {author} {\bibfnamefont {M.}~\bibnamefont {Fatyga}}, \bibinfo {author} {\bibfnamefont {K.}~\bibnamefont {Kwiatkowski}}, \bibinfo {author} {\bibfnamefont {V.~E.}\ \bibnamefont {Viola}}, \bibinfo {author} {\bibfnamefont {C.~B.}\ \bibnamefont {Chitwood}}, \bibinfo {author} {\bibfnamefont {D.~J.}\ \bibnamefont {Fields}}, \bibinfo {author} {\bibfnamefont {C.~K.}\ \bibnamefont {Gelbke}}, \bibinfo {author} {\bibfnamefont {W.~G.}\ \bibnamefont {Lynch}}, \bibinfo {author} {\bibfnamefont {J.}~\bibnamefont {Pochodzalla}}, \bibinfo {author} {\bibfnamefont {M.~B.}\ \bibnamefont {Tsang}},\ and\ \bibinfo {author} {\bibfnamefont {M.}~\bibnamefont {Blann}},\ }\href {https://doi.org/10.1103/PhysRevLett.55.1376} {\bibfield  {journal} {\bibinfo  {journal} {Phys. Rev. Lett.}\ }\textbf {\bibinfo {volume} {55}},\ \bibinfo {pages} {1376} (\bibinfo {year} {1985})}\BibitemShut {NoStop}%
\bibitem [{\citenamefont {Fuchs}\ and\ \citenamefont {Mohring}(1994)}]{Fuchs1994RPP}%
  \BibitemOpen
\bibfield  {author} {\bibinfo {author} {\bibfnamefont {H.}~\bibnamefont {Fuchs}}\ and\ \bibinfo {author} {\bibfnamefont {K.}~\bibnamefont {M{\"o}hring}},\ }\href {https://doi.org/10.1088/0034-4885/57/3/001} {\bibfield  {journal} {\bibinfo  {journal} {Rep. Prog. Phys.}\ }\textbf {\bibinfo {volume} {57}},\ \bibinfo {pages} {231} (\bibinfo {year} {1994})}\BibitemShut {NoStop}%
\bibitem [{\citenamefont {{Particle Therapy Co-operative Group (PTCOG)}}(2025)}]{web_PTCOG}%
  \BibitemOpen
  \bibfield  {author} {\bibinfo {author} {\bibnamefont {{Particle Therapy Co-operative Group (PTCOG)}}},\ }\href@noop {} {\bibinfo {title} {Particle therapy facilities in clinical operation}},\ \bibinfo {howpublished} {\url{https://www.ptcog.site/index.php/facilities-in-operation-public}} (\bibinfo {year} {2025}),\ \bibinfo {note} {accessed: 28 November 2025}\BibitemShut {NoStop}%
\bibitem [{\citenamefont {Mohamad}\ \emph {et~al.}(2017)\citenamefont {Mohamad}, \citenamefont {Sishc}, \citenamefont {Saha}, \citenamefont {Pompos}, \citenamefont {Rahimi}, \citenamefont {Story}, \citenamefont {Davis},\ and\ \citenamefont {Kim}}]{Mohamad2017Cancers}%
  \BibitemOpen
  \bibfield  {author} {\bibinfo {author} {\bibfnamefont {O.}~\bibnamefont {Mohamad}}, \bibinfo {author} {\bibfnamefont {B.~J.}\ \bibnamefont {Sishc}}, \bibinfo {author} {\bibfnamefont {J.}~\bibnamefont {Saha}}, \bibinfo {author} {\bibfnamefont {A.}~\bibnamefont {Pompos}}, \bibinfo {author} {\bibfnamefont {A.}~\bibnamefont {Rahimi}}, \bibinfo {author} {\bibfnamefont {M.~D.}\ \bibnamefont {Story}}, \bibinfo {author} {\bibfnamefont {A.~J.}\ \bibnamefont {Davis}},\ and\ \bibinfo {author} {\bibfnamefont {D.~W.~N.}\ \bibnamefont {Kim}},\ }\href {https://doi.org/10.3390/cancers9060066} {\bibfield  {journal} {\bibinfo  {journal} {Cancers}\ }\textbf {\bibinfo {volume} {9}},\ \bibinfo {pages} {66} (\bibinfo {year} {2017})}\BibitemShut {NoStop}%
\bibitem [{\citenamefont {Kostyleva}\ \emph {et~al.}(2023)\citenamefont {Kostyleva} \emph {et~al.}}]{Kostyleva2023PMB}%
  \BibitemOpen
  \bibfield  {author} {\bibinfo {author} {\bibfnamefont {D.}~\bibnamefont {Kostyleva}} \emph {et~al.} (\bibinfo {collaboration} {Super-FRS Experiment Collaboration}),\ }\href {https://doi.org/10.1088/1361-6560/aca5e8} {\bibfield  {journal} {\bibinfo  {journal} {Phys. Med. Biol.}\ }\textbf {\bibinfo {volume} {68}},\ \bibinfo {pages} {015003} (\bibinfo {year} {2023})}\BibitemShut {NoStop}%
\bibitem [{\citenamefont {Nandy}(2021)}]{Nandy2021FP}%
  \BibitemOpen
  \bibfield  {author} {\bibinfo {author} {\bibfnamefont {M.}~\bibnamefont {Nandy}},\ }\href {https://doi.org/10.3389/fphy.2020.598257} {\bibfield  {journal} {\bibinfo  {journal} {Front. Phys.}\ }\textbf {\bibinfo {volume} {8}},\ \bibinfo {pages} {598257} (\bibinfo {year} {2021})}\BibitemShut {NoStop}%
\bibitem [{\citenamefont {Dudouet}\ \emph {et~al.}(2014)\citenamefont {Dudouet}, \citenamefont {Labalme}, \citenamefont {Cussol}, \citenamefont {Finck}, \citenamefont {Rescigno}, \citenamefont {Rousseau}, \citenamefont {Salvador},\ and\ \citenamefont {Vanstalle}}]{Dudouet2014PRC}%
  \BibitemOpen
  \bibfield  {author} {\bibinfo {author} {\bibfnamefont {J.}~\bibnamefont {Dudouet}}, \bibinfo {author} {\bibfnamefont {M.}~\bibnamefont {Labalme}}, \bibinfo {author} {\bibfnamefont {D.}~\bibnamefont {Cussol}}, \bibinfo {author} {\bibfnamefont {C.}~\bibnamefont {Finck}}, \bibinfo {author} {\bibfnamefont {R.}~\bibnamefont {Rescigno}}, \bibinfo {author} {\bibfnamefont {M.}~\bibnamefont {Rousseau}}, \bibinfo {author} {\bibfnamefont {S.}~\bibnamefont {Salvador}},\ and\ \bibinfo {author} {\bibfnamefont {M.}~\bibnamefont {Vanstalle}},\ }\href {https://doi.org/10.1103/PhysRevC.89.064615} {\bibfield  {journal} {\bibinfo  {journal} {Phys. Rev. C}\ }\textbf {\bibinfo {volume} {89}},\ \bibinfo {pages} {064615} (\bibinfo {year} {2014})}\BibitemShut {NoStop}%
\bibitem [{\citenamefont {Divay}\ \emph {et~al.}(2017)\citenamefont {Divay}, \citenamefont {Colin}, \citenamefont {Cussol}, \citenamefont {Finck}, \citenamefont {Karakaya}, \citenamefont {Labalme}, \citenamefont {Rousseau}, \citenamefont {Salvador},\ and\ \citenamefont {Vanstalle}}]{Divay2017PRC}%
  \BibitemOpen
  \bibfield  {author} {\bibinfo {author} {\bibfnamefont {C.}~\bibnamefont {Divay}}, \bibinfo {author} {\bibfnamefont {J.}~\bibnamefont {Colin}}, \bibinfo {author} {\bibfnamefont {D.}~\bibnamefont {Cussol}}, \bibinfo {author} {\bibfnamefont {C.}~\bibnamefont {Finck}}, \bibinfo {author} {\bibfnamefont {Y.}~\bibnamefont {Karakaya}}, \bibinfo {author} {\bibfnamefont {M.}~\bibnamefont {Labalme}}, \bibinfo {author} {\bibfnamefont {M.}~\bibnamefont {Rousseau}}, \bibinfo {author} {\bibfnamefont {S.}~\bibnamefont {Salvador}},\ and\ \bibinfo {author} {\bibfnamefont {M.}~\bibnamefont {Vanstalle}},\ }\href {https://doi.org/10.1103/PhysRevC.95.044602} {\bibfield  {journal} {\bibinfo  {journal} {Phys. Rev. C}\ }\textbf {\bibinfo {volume} {95}},\ \bibinfo {pages} {044602} (\bibinfo {year} {2017})}\BibitemShut {NoStop}%
\bibitem [{\citenamefont {De~Napoli}\ \emph {et~al.}(2012)\citenamefont {De~Napoli} \emph {et~al.}}]{Napoli2012PMB}%
  \BibitemOpen
  \bibfield  {author} {\bibinfo {author} {\bibfnamefont {M.}~\bibnamefont {De~Napoli}} \emph {et~al.},\ }\href {https://doi.org/10.1088/0031-9155/57/22/7651} {\bibfield  {journal} {\bibinfo  {journal} {Phys. Med. Biol.}\ }\textbf {\bibinfo {volume} {57}},\ \bibinfo {pages} {7651} (\bibinfo {year} {2012})}\BibitemShut {NoStop}%
\bibitem [{\citenamefont {Bougault}\ \emph {et~al.}(2014)\citenamefont {Bougault} \emph {et~al.}}]{Bouga2014EPJA}%
  \BibitemOpen
  \bibfield  {author} {\bibinfo {author} {\bibfnamefont {R.}~\bibnamefont {Bougault}} \emph {et~al.} (\bibinfo {collaboration} {FAZIA Collaboration}),\ }\href {https://doi.org/10.1140/epja/i2014-14047-4} {\bibfield  {journal} {\bibinfo  {journal} {Eur. Phys. J. A}\ }\textbf {\bibinfo {volume} {50}},\ \bibinfo {pages} {47} (\bibinfo {year} {2014})}\BibitemShut {NoStop}%
\bibitem [{\citenamefont {Valdré}\ \emph {et~al.}(2019)\citenamefont {Valdré} \emph {et~al.}}]{Valdre2019NIMA}%
  \BibitemOpen
  \bibfield  {author} {\bibinfo {author} {\bibfnamefont {S.}~\bibnamefont {Valdré}} \emph {et~al.},\ }\href {https://doi.org/10.1016/j.nima.2019.03.082} {\bibfield  {journal} {\bibinfo  {journal} {Nucl. Instrum. Methods Phys. Res. A}\ }\textbf {\bibinfo {volume} {930}},\ \bibinfo {pages} {27--36} (\bibinfo {year} {2019})}\BibitemShut {NoStop}%
\bibitem [{\citenamefont {Frosin}\ \emph {et~al.}(2020)\citenamefont {Frosin} \emph {et~al.}}]{Frosin2020NIMA}%
  \BibitemOpen
 \bibfield  {author} {\bibinfo {author} {\bibfnamefont {C.}~\bibnamefont {Frosin}} \emph {et~al.},\ }\href {https://doi.org/10.1016/j.nima.2019.163018} {\bibfield  {journal} {\bibinfo  {journal} {Nucl. Instrum. Methods Phys. Res. A}\ }\textbf
  {\bibinfo {volume} {951}},\ \bibinfo {pages} {163018} (\bibinfo {year} {2020})}\BibitemShut {NoStop}%
\bibitem [{\citenamefont {Baldesi}\ \emph {et~al.}(2025)\citenamefont {Baldesi} \emph {et~al.}}]{Baldesi2025NIMA}%
  \BibitemOpen
\bibfield  {author} {\bibinfo {author} {\bibfnamefont {L.}~\bibnamefont {Baldesi}} \emph {et~al.},\ }\href {https://doi.org/10.1016/j.nima.2025.170420} {\bibfield  {journal} {\bibinfo  {journal} {Nucl. Instrum. Methods Phys. Res. A}\ }\textbf
  {\bibinfo {volume} {1075}},\ \bibinfo {pages} {170420} (\bibinfo {year} {2025})}\BibitemShut {NoStop}%
\bibitem [{\citenamefont {Baldesi}\ \emph {et~al.}(2026)\citenamefont {Baldesi} \emph {et~al.}}]{Baldesi2026NIMA}%
  \BibitemOpen
  \bibfield  {author} {\bibinfo {author} {\bibfnamefont {L.}~\bibnamefont {Baldesi}} \emph {et~al.},\ }\href {https://doi.org/10.1016/j.nima.2026.171865} {\bibfield  {journal} {\bibinfo  {journal} {Nucl. Instrum. Methods Phys. Res. A}\ }\textbf {\bibinfo {volume} {1092}},\ \bibinfo {pages} {171865} (\bibinfo {year} {2026})}\BibitemShut {NoStop}%
\bibitem [{\citenamefont {Frankland}\ \emph {et~al.}(2002)\citenamefont {Frankland}, \citenamefont {Bonnet},\ and\ \citenamefont {Gruyer}}]{kaliveda}%
  \BibitemOpen
  \bibfield  {author} {\bibinfo {author} {\bibfnamefont {J.~D.}~\bibnamefont {Frankland}}, \bibinfo {author} {\bibfnamefont {E.}~\bibnamefont {Bonnet}},\ and\ \bibinfo {author} {\bibfnamefont {D.}~\bibnamefont {Gruyer}},\ }\href {https://doi.org/10.5281/zenodo.7664901} {\bibinfo {title} {Kaliveda: Heavy-ion collisions analysis toolkit}} (\bibinfo {year} {2002})\BibitemShut {NoStop}%
\bibitem [{\citenamefont {Pastore}\ \emph {et~al.}(2017)\citenamefont {Pastore} \emph {et~al.}}]{Pastore2017NIMA}%
  \BibitemOpen
  \bibfield  {author} {\bibinfo {author} {\bibfnamefont {G.}~\bibnamefont {Pastore}} \emph {et~al.} (\bibinfo {collaboration} {FAZIA Collaboration}),\ }\href {https://doi.org/10.1016/j.nima.2017.01.048} {\bibfield  {journal} {\bibinfo  {journal} {Nucl. Instrum. Methods Phys. Res. A}\ }\textbf {\bibinfo {volume} {860}},\ \bibinfo {pages} {42--50} (\bibinfo {year} {2017})}\BibitemShut {NoStop}%
  \bibitem [{\citenamefont {Rebillard-Souli{\'e}}(2024)}]{Alex2024thesis}%
  \BibitemOpen
  \bibfield  {author} {\bibinfo {author} {\bibfnamefont {A.}~\bibnamefont {Rebillard-Souli{\'e}}},\ }\href@noop {} {\bibinfo {type} {{Ph.D. thesis}}},\ \bibinfo  {school} {University of Caen Normandie} (\bibinfo {year} {2024})\BibitemShut {NoStop}%
\bibitem [{\citenamefont {Pârlog}\ \emph {et~al.}(2002)\citenamefont {Pârlog} \emph {et~al.}}]{Parlog2002NIMA}%
  \BibitemOpen
  \bibfield  {author} {\bibinfo {author} {\bibfnamefont {M.}~\bibnamefont {Pârlog}} \emph {et~al.},\ }\href {https://doi.org/10.1016/S0168-9002(01)01710-7} {\bibfield  {journal} {\bibinfo  {journal} {Nucl. Instrum. Methods Phys. Res. A}\ }\textbf {\bibinfo {volume} {482}},\ \bibinfo {pages} {674--692} (\bibinfo {year} {2002})}\BibitemShut {NoStop}%
\bibitem [{\citenamefont {Morfouace}\ \emph {et~al.}(2017)\citenamefont {Morfouace}, \citenamefont {Lynch},\ and\ \citenamefont {Tsang}}]{morfo2017NIMA}%
  \BibitemOpen
\bibfield  {author} {\bibinfo {author} {\bibfnamefont {P.}~\bibnamefont {Morfouace}}, \bibinfo {author} {\bibfnamefont
  {W.~G.}~\bibnamefont {Lynch}},\ and\ \bibinfo {author} {\bibfnamefont {M.~B.}~\bibnamefont {Tsang}},\ }\href {https://doi.org/10.1016/j.nima.2016.12.045} {\bibfield  {journal} {\bibinfo  {journal} {Nucl. Instrum. Methods Phys. Res. A}\ }\textbf {\bibinfo {volume} {848}},\ \bibinfo {pages} {45--53} (\bibinfo {year} {2017})}\BibitemShut {NoStop}%
\bibitem [{\citenamefont {Sweany}\ \emph {et~al.}(2021)\citenamefont {Sweany}, \citenamefont {Lynch}, \citenamefont {Brown}, \citenamefont {Anthony}, \citenamefont {Chajecki}, \citenamefont {Dell’Aquila}, \citenamefont {Morfouace}, \citenamefont {Teh}, \citenamefont {Tsang}, \citenamefont {Tsang}, \citenamefont {Wang}\ and\ \citenamefont {Zhu}}]{Swean2021NIMA}%
  \BibitemOpen
\bibfield  {author} {\bibinfo {author} {\bibfnamefont {S.}~\bibnamefont {Sweany}}, \bibinfo {author} {\bibfnamefont {W.~G.}~\bibnamefont {Lynch}}, \bibinfo {author} {\bibfnamefont {K.}~\bibnamefont {Brown}}, \bibinfo {author} {\bibfnamefont {A.}~\bibnamefont {Anthony}}, \bibinfo {author} {\bibfnamefont {Z.}~\bibnamefont {Chajecki}}, \bibinfo {author} {\bibfnamefont {D.}~\bibnamefont {Dell’Aquila}}, \bibinfo {author} {\bibfnamefont {P.}~\bibnamefont {Morfouace}}, \bibinfo {author} {\bibfnamefont {F.~C.~E.}~\bibnamefont {Teh}}, \bibinfo {author} {\bibfnamefont {C.~Y.}~\bibnamefont {Tsang}}, \bibinfo {author} {\bibfnamefont {M.~B.}~\bibnamefont {Tsang}}, \bibinfo {author} {\bibfnamefont {R.~S.}~\bibnamefont {Wang}},\ and\ \bibinfo {author} {\bibfnamefont {K.}~\bibnamefont {Zhu}},\ }\href {https://doi.org/10.1016/j.nima.2021.165798} {\bibfield  {journal} {\bibinfo  {journal} {Nucl. Instrum. Methods Phys. Res. A}\ }\textbf {\bibinfo {volume} {1018}},\ \bibinfo {pages} {165798} (\bibinfo {year} {2021})}\BibitemShut {NoStop}%
\bibitem [{\citenamefont {Thompson}(1988)}]{Thompson1988CPR}%
  \BibitemOpen
  \bibfield  {author} {\bibinfo {author} {\bibfnamefont {Ian~J.}\ \bibnamefont {Thompson}},\ }\href {https://doi.org/10.1016/0167-7977(88)90005-6} {\bibfield  {journal} {\bibinfo  {journal} {Comput. Phys. Rep.}\ }\textbf {\bibinfo {volume} {7}},\ \bibinfo {pages} {167--212} (\bibinfo {year} {1988})}\BibitemShut {NoStop}%
\bibitem [{\citenamefont {Buenerd}\ \emph {et~al.}(1984)\citenamefont {Buenerd}, \citenamefont {Lounis}, \citenamefont {Chauvin}, \citenamefont {Lebrun}, \citenamefont {Martin}, \citenamefont {Duhamel}, \citenamefont {Gondrand},\ and\ \citenamefont {{De Saintignon}}}]{Buenerd1984NPA}%
  \BibitemOpen
  \bibfield  {author} {\bibinfo {author} {\bibfnamefont {M.}~\bibnamefont {Buenerd}}, \bibinfo {author} {\bibfnamefont {A.}~\bibnamefont {Lounis}}, \bibinfo {author} {\bibfnamefont {J.}~\bibnamefont {Chauvin}}, \bibinfo {author} {\bibfnamefont {D.}~\bibnamefont {Lebrun}}, \bibinfo {author} {\bibfnamefont {P.}~\bibnamefont {Martin}}, \bibinfo {author} {\bibfnamefont {G.}~\bibnamefont {Duhamel}}, \bibinfo {author} {\bibfnamefont {J.~C.}~\bibnamefont {Gondrand}},\ and\ \bibinfo {author} {\bibfnamefont {P.}~\bibnamefont {{De Saintignon}}},\ }\href {https://doi.org/10.1016/0375-9474(84)90186-6} {\bibfield  {journal} {\bibinfo  {journal} {Nucl. Phys. A}\ }\textbf {\bibinfo {volume} {424}},\ \bibinfo {pages} {313--334} (\bibinfo {year} {1984})}\BibitemShut {NoStop}%
\bibitem [{\citenamefont {Goldhaber}(1974)}]{Goldhaber1974PLB}%
  \BibitemOpen
  \bibfield  {author} {\bibinfo {author} {\bibfnamefont {A.~S.}\ \bibnamefont {Goldhaber}},\ }\href {https://doi.org/10.1016/0370-2693(74)90388-8} {\bibfield  {journal} {\bibinfo  {journal} {Phys. Lett. B}\ }\textbf {\bibinfo {volume} {53}},\ \bibinfo {pages} {306--308} (\bibinfo {year} {1974})}\BibitemShut {NoStop}%
\bibitem [{\citenamefont {Giacomelli}\ \emph {et~al.}(2004)\citenamefont {Giacomelli}, \citenamefont {Sihver}, \citenamefont {Skvar\ifmmode~\check{c}\else \v{c}\fi{}}, \citenamefont {Yasuda},\ and\ \citenamefont {Ili\ifmmode~\acute{c}\else \'{c}\fi{}}}]{Giacomelli2004PRC}%
  \BibitemOpen
  \bibfield  {author} {\bibinfo {author} {\bibfnamefont {M.}~\bibnamefont {Giacomelli}}, \bibinfo {author} {\bibfnamefont {L.}~\bibnamefont {Sihver}}, \bibinfo {author} {\bibfnamefont {J.}~\bibnamefont {Skvar\ifmmode~\check{c}\else \v{c}\fi{}}}, \bibinfo {author} {\bibfnamefont {N.}~\bibnamefont {Yasuda}},\ and\ \bibinfo {author} {\bibfnamefont {R.}~\bibnamefont {Ili\ifmmode~\acute{c}\else \'{c}\fi{}}},\ }\href {https://doi.org/10.1103/PhysRevC.69.064601} {\bibfield  {journal} {\bibinfo  {journal} {Phys. Rev. C}\ }\textbf {\bibinfo {volume} {69}},\ \bibinfo {pages} {064601} (\bibinfo {year} {2004})}\BibitemShut {NoStop}%
\bibitem [{\citenamefont {Van~Bibber}\ \emph {et~al.}(1979)\citenamefont {Van~Bibber}, \citenamefont {Hendrie}, \citenamefont {Scott}, \citenamefont {Weiman}, \citenamefont {Schroeder}, \citenamefont {Geaga}, \citenamefont {Cessin}, \citenamefont {Treuhaft}, \citenamefont {Grossiord}, \citenamefont {Rasmussen},\ and\ \citenamefont {Wong}}]{VanBibber1979PRL}%
  \BibitemOpen
  \bibfield  {author} {\bibinfo {author} {\bibfnamefont {K.}~\bibnamefont {Van~Bibber}}, \bibinfo {author} {\bibfnamefont {D.~L.}\ \bibnamefont {Hendrie}}, \bibinfo {author} {\bibfnamefont {D.~K.}\ \bibnamefont {Scott}}, \bibinfo {author} {\bibfnamefont {H.~H.}\ \bibnamefont {Weiman}}, \bibinfo {author} {\bibfnamefont {L.~S.}\ \bibnamefont {Schroeder}}, \bibinfo {author} {\bibfnamefont {J.~V.}\ \bibnamefont {Geaga}}, \bibinfo {author} {\bibfnamefont {S.~A.}\ \bibnamefont {Cessin}}, \bibinfo {author} {\bibfnamefont {R.}~\bibnamefont {Treuhaft}}, \bibinfo {author} {\bibfnamefont {Y.~J.}\ \bibnamefont {Grossiord}}, \bibinfo {author} {\bibfnamefont {J.~O.}\ \bibnamefont {Rasmussen}},\ and\ \bibinfo {author} {\bibfnamefont {C.~Y.}\ \bibnamefont {Wong}},\ }\href {https://doi.org/10.1103/PhysRevLett.43.840} {\bibfield  {journal} {\bibinfo  {journal} {Phys. Rev. Lett.}\ }\textbf {\bibinfo {volume} {43}},\ \bibinfo {pages} {840} (\bibinfo {year} {1979})}\BibitemShut {NoStop}%
\bibitem [{\citenamefont {Notani}\ \emph {et~al.}(2007)\citenamefont {Notani} \emph {et~al.}}]{Notani2007PRC}%
  \BibitemOpen
  \bibfield  {author} {\bibinfo {author} {\bibfnamefont {M.}~\bibnamefont {Notani}} \emph {et~al.},\ }\href {https://doi.org/10.1103/PhysRevC.76.044605} {\bibfield  {journal} {\bibinfo  {journal} {Phys. Rev. C}\ }\textbf {\bibinfo {volume} {76}},\ \bibinfo {pages} {044605} (\bibinfo {year} {2007})}\BibitemShut {NoStop}%
\bibitem [{\citenamefont {Momota}\ \emph {et~al.}(2023)\citenamefont {Momota}, \citenamefont {Ohtsubo}, \citenamefont {Honma}, \citenamefont {Kitagawa},\ and\ \citenamefont {Sato}}]{Momota2023PhysScr}%
  \BibitemOpen
\bibfield  {author} {\bibinfo {author} {\bibfnamefont {S.}~\bibnamefont {Momota}}, \bibinfo {author} {\bibfnamefont {T.}~\bibnamefont {Ohtsubo}}, \bibinfo {author} {\bibfnamefont {A.}~\bibnamefont {Honma}}, \bibinfo {author} {\bibfnamefont {A.}~\bibnamefont {Kitagawa}},\ and\ \bibinfo {author} {\bibfnamefont {S.}~\bibnamefont {Sato}},\ }\href {https://doi.org/10.1088/1402-4896/acdf93} {\bibfield  {journal} {\bibinfo  {journal} {Phys. Scr.}\ }\textbf {\bibinfo {volume} {98}},\ \bibinfo {pages} {085301} (\bibinfo {year} {2023})}\BibitemShut {NoStop}%
\bibitem [{\citenamefont {Wong}\ and\ \citenamefont {Van~Bibber}(1982)}]{wong1982PRC}%
  \BibitemOpen
  \bibfield  {author} {\bibinfo {author} {\bibfnamefont {C.~Y.}\ \bibnamefont {Wong}}\ and\ \bibinfo {author} {\bibfnamefont {K.}~\bibnamefont {Van~Bibber}},\ }\href {https://doi.org/10.1103/PhysRevC.25.2990} {\bibfield  {journal} {\bibinfo  {journal} {Phys. Rev. C}\ }\textbf {\bibinfo {volume} {25}},\ \bibinfo {pages} {2990} (\bibinfo {year} {1982})}\BibitemShut {NoStop}%
\bibitem [{\citenamefont {Greiner}\ \emph {et~al.}(1975)\citenamefont {Greiner}, \citenamefont {Lindstrom}, \citenamefont {Heckman}, \citenamefont {Cork},\ and\ \citenamefont {Bieser}}]{Greiner1975PRL}%
  \BibitemOpen
  \bibfield  {author} {\bibinfo {author} {\bibfnamefont {D.~E.}\ \bibnamefont {Greiner}}, \bibinfo {author} {\bibfnamefont {P.~J.}\ \bibnamefont {Lindstrom}}, \bibinfo {author} {\bibfnamefont {H.~H.}\ \bibnamefont {Heckman}}, \bibinfo {author} {\bibfnamefont {B.}~\bibnamefont {Cork}},\ and\ \bibinfo {author} {\bibfnamefont {F.~S.}\ \bibnamefont {Bieser}},\ }\href {https://doi.org/10.1103/PhysRevLett.35.152} {\bibfield  {journal} {\bibinfo  {journal} {Phys. Rev. Lett.}\ }\textbf {\bibinfo {volume} {35}},\ \bibinfo {pages} {152} (\bibinfo {year} {1975})}\BibitemShut {NoStop}%
\bibitem [{\citenamefont {Dudouet}\ \emph {et~al.}(2013)\citenamefont {Dudouet}, \citenamefont {Juliani}, \citenamefont {Labalme}, \citenamefont {Cussol}, \citenamefont {Angelique}, \citenamefont {Braunn}, \citenamefont {Colin}, \citenamefont {Finck}, \citenamefont {Fontbonne}, \citenamefont {Guerin}, \citenamefont {Henriquet}, \citenamefont {Krimmer}, \citenamefont {Rousseau}, \citenamefont {Saint-Laurent},\ and\ \citenamefont {Salvador}}]{Dudouet2013PRC}%
  \BibitemOpen
  \bibfield  {author} {\bibinfo {author} {\bibfnamefont {J.}~\bibnamefont {Dudouet}}, \bibinfo {author} {\bibfnamefont {D.}~\bibnamefont {Juliani}}, \bibinfo {author} {\bibfnamefont {M.}~\bibnamefont {Labalme}}, \bibinfo {author} {\bibfnamefont {D.}~\bibnamefont {Cussol}}, \bibinfo {author} {\bibfnamefont {J.~C.}\ \bibnamefont {Ang{\'e}lique}}, \bibinfo {author} {\bibfnamefont {B.}~\bibnamefont {Braunn}}, \bibinfo {author} {\bibfnamefont {J.}~\bibnamefont {Colin}}, \bibinfo {author} {\bibfnamefont {C.}~\bibnamefont {Finck}}, \bibinfo {author} {\bibfnamefont {J.~M.}\ \bibnamefont {Fontbonne}}, \bibinfo {author} {\bibfnamefont {H.}~\bibnamefont {Gu{\'e}rin}}, \bibinfo {author} {\bibfnamefont {P.}~\bibnamefont {Henriquet}}, \bibinfo {author} {\bibfnamefont {J.}~\bibnamefont {Krimmer}}, \bibinfo {author} {\bibfnamefont {M.}~\bibnamefont {Rousseau}}, \bibinfo {author} {\bibfnamefont {M.~G.}\ \bibnamefont {Saint-Laurent}},\ and\ \bibinfo {author} {\bibfnamefont {S.}~\bibnamefont {Salvador}},\ }\href
  {https://doi.org/10.1103/PhysRevC.88.024606} {\bibfield  {journal} {\bibinfo  {journal} {Phys. Rev. C}\ }\textbf {\bibinfo {volume} {88}},\ \bibinfo {pages} {024606} (\bibinfo {year} {2013})}\BibitemShut {NoStop}%
\end{thebibliography}
\end{document}